\documentclass[conference]{IEEEtran}
\IEEEoverridecommandlockouts
\usepackage{cite}
\usepackage{amsmath,amssymb,amsfonts}
\usepackage{algorithmic}
\usepackage{graphicx}
\usepackage{textcomp}
\usepackage{xcolor}
\def\BibTeX{{\rm B\kern-.05em{\sc i\kern-.025em b}\kern-.08em
    T\kern-.1667em\lower.7ex\hbox{E}\kern-.125emX}}
\begin{document}

\title{Exploration of LLM Multi-Agent Application Implementation Based on LangGraph+CrewAI\\
 
}

\author{\IEEEauthorblockN{Zhihua Duan}
\IEEEauthorblockA{\textit{Intelligent Cloud Network Monitoring Department  } \\
\textit{China Telecom Shanghai Company}\\
Shanghai, China \\
duanzh.sh@chinatelecom.cn}
\and
\IEEEauthorblockN{Jialin Wang}
\IEEEauthorblockA{\textit{Executive Vice President  } \\
\textit{Ferret Relationship Intelligence }\\
Burlingame, CA 94010, USA \\
jialinwangspace@gmail.com} 
}

\maketitle

\begin{abstract}
With the rapid development of large model technology, the application of agent technology in various fields is becoming increasingly widespread, profoundly changing people's work and lifestyles. In complex and dynamic systems, multi-agents achieve complex tasks that are difficult for a single agent to complete through division of labor and collaboration among agents. This paper discusses the integrated application of LangGraph and CrewAI. LangGraph improves the efficiency of information transmission through graph architecture, while CrewAI enhances team collaboration capabilities and system performance through intelligent task allocation and resource management. The main research contents of this paper are: (1) designing the architecture of agents based on LangGraph for precise control; (2) enhancing the capabilities of agents based on CrewAI to complete a variety of tasks. This study aims to delve into the application of LangGraph and CrewAI in multi-agent systems, providing new perspectives for the future development of agent technology, and promoting technological progress and application innovation in the field of large model intelligent agents.

\end{abstract}

\begin{IEEEkeywords}
LangGraph, CrewAI, Agent, Large Language Model, Generative AI
\end{IEEEkeywords}

\section{Introduction}
Driven by the rapid advancements in artificial intelligence technology, the application of agents in various fields is increasingly growing, profoundly affecting everyone's work and lifestyle. The application scope of agent technology is broad; it can autonomously perceive the environment, conduct data analysis, and make decisions, thereby significantly enhancing efficiency and optimizing resource allocation. In complex and dynamic systems, the introduction of multi-agent systems enables multiple agents to collaborate and complete complex tasks that are difficult for a single agent to achieve.

The key advantage of multi-agent systems lies in their task decomposition capabilities, achieving goals through the collaborative action of agents, which not only enhances the system's flexibility and adaptability but also improves its generalization ability. In environments characterized by uncertainty and dynamic changes, the ability of agents to collaborate and divide tasks is particularly crucial.

This paper primarily investigates two issues: (1) designing the architecture of agents based on LangGraph for more precise control; (2) enhancing the capabilities of agents based on CrewAI to accomplish different tasks. The aim is to delve into the application of AI multi-agent systems, explore the technological advantages brought about by the combination of LangGraph and CrewAI, and their potential application value in various fields. Through the analysis and research of these technologies, it is expected to provide new perspectives and ideas for the future development of agent technology, thereby promoting technological progress and application innovation in the field of large model intelligent agents.

\section{Related Work}
Autonomous agents are typically responsible for specific roles to accomplish various tasks.MetaGPT\cite{MetaGPT}, ChatDev\cite{Chatdev}, and self-collaboration\cite{Self-collaboration} pre-set various roles and corresponding responsibilities to foster collaboration among agents.

Autonomous agents based on Large Language Models (LLMs) incorporate mechanisms from human memory processes.RLP\cite{RLP} is a conversational agent that maintains an internal state for both parties in a dialogue, achieving the agent's short-term memory.SayPlan\cite{Sayplan} is an agent designed for task planning and design.When faced with complex tasks, break them down into simpler subtasks and solve them. The Chain of Thought (CoT) \cite{CoT} inputs reasoning steps for solving complex problems into the prompt.\cite{LangGraph} proposes an intelligent personalized digital banking assistant based on LangGraph and Chain of Thoughts (COT), leveraging Large Language Models (LLM) and a multi-agent framework to enhance task efficiency.
\cite{Graph} improves advanced question-answering systems based on Retrieval-Augmented Generation (RAG) by leveraging graph technology, to overcome the limitations of existing RAG models and develop high-quality artificial intelligence services.

In this study, autonomous agents based on LLM can leverage the framework capabilities of LangGraph and CrewAI to automatically perform various tasks, endowing Agents with the ability to complete specific tasks, forming a comprehensive application system framework.

\section{Methodology}

\subsection{Introduction to LangGraph} 
LangGraph is a framework designed for constructing multi-agent applications, enabling developers to utilize large language models (LLMs) to create agents and multi-agent workflows. Compared to other LLM frameworks, LangGraph offers the benefits of loops, controllability, and persistent memory. LangGraph provides granular control over the application's processes and states, enabling the creation of reliable agents that support advanced human-computer interaction and memory capabilities. The flexible framework of LangGraph supports various control flows—single-agent, multi-agent, hierarchical, sequential and can robustly manage complex real-world scenarios. 

\subsection{CrewAI Framework}
CrewAI is an open-source framework designed to coordinate AI agents with role-playing and autonomous operations to facilitate cooperation among agents in solving complex problems. It allows developers to define AI agents with specific roles, objectives, and tools. The main building blocks of CrewAI include Agent, Task, Tool, and Crew, offering a rich set of features that can be freely selected and combined according to specific needs to create multi-agent systems. CrewAI supports various APIs such as OpenAI and Ollama, and it has key features like role-customized agents, automatic task delegation, and flexibility in task management. CrewAI is aimed at handling complex tasks, such as multi-step workflows, decision-making, and problem-solving.

An Agent intelligent entity is an autonomous unit capable of performing tasks, making decisions, and communicating with other agents. In the CrewAI framework, a Task refers to the specific work carried out by an Agent, including details such as description, executing agents, and required tools. It supports multi-agent collaboration and optimizes team cooperation and efficiency through the process orchestration of the Crew.

\begin{figure}[htbp]
  \centering
  \includegraphics[width=0.6\linewidth]{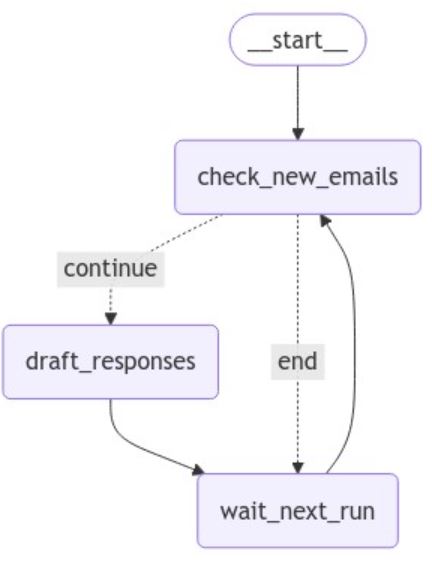}
  \caption{Illustration of the email case based on LangGraph +CrewAI.}
\end{figure}

\subsection{Integrating LangGraph with CrewAI}
The integration of LangGraph+CrewAI framework provides powerful tools for complex task management, optimizing task execution and multi-agent systems through flexible workflows, inter-agent collaboration, and graph-structured management. It supports customized development by integrating with existing tools, meeting the evolving needs of AI applications. CrewAI ensures process efficiency through clear task allocation and role definition. Additionally, seamless integration with LangChain allows developers familiar with the framework to easily integrate independent agents, further enhancing the framework's appeal.

Taking the example of the automatic email composition and sending case provided on the CrewAI official website, the LangGraph+CrewAI framework breaks down complex tasks into manageable steps and automates their execution.

As shown in Figure 1, with LangGraph, it is possible to clearly define and visualize the workflow for writing and sending emails, including checking new emails, composing new emails, and waiting for the next run.

As shown in Figures 2 to 4 ,this is a code example  based on CrewAI+LangGraph. By integrating CrewAI with LangGraph, it becomes easy to implement functionalities such as email checking, composing, and automatic sending based on large models.

\begin{figure}[htbp]
  \centering
  \includegraphics[width=\linewidth]{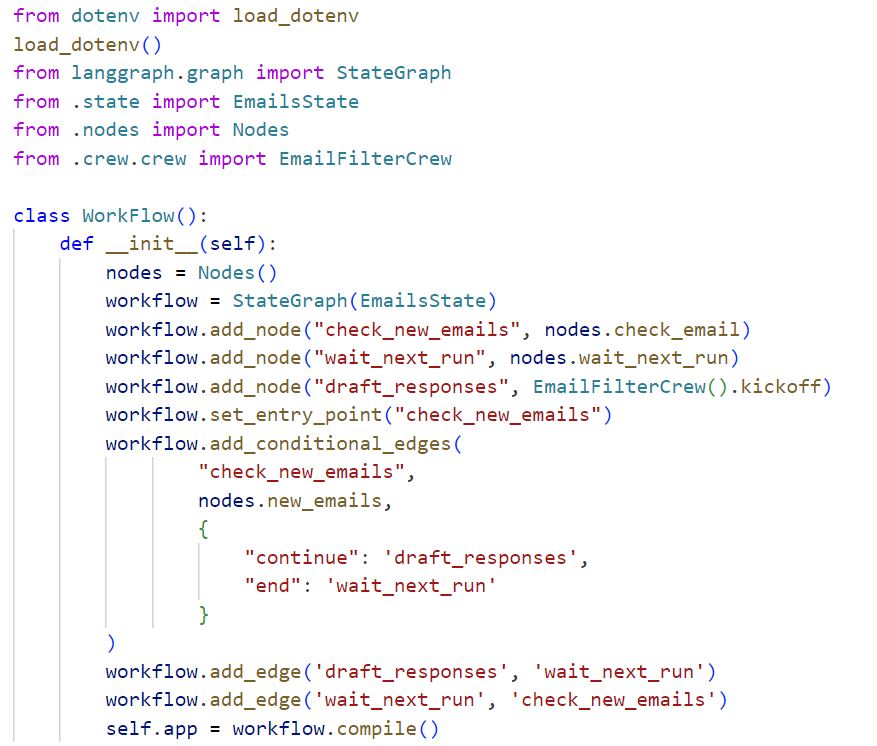}
  \caption{Code example based on LangGraph + CrewAI.
}
\end{figure}

\begin{figure}[htbp]
  \centering
  \includegraphics[width=\linewidth]{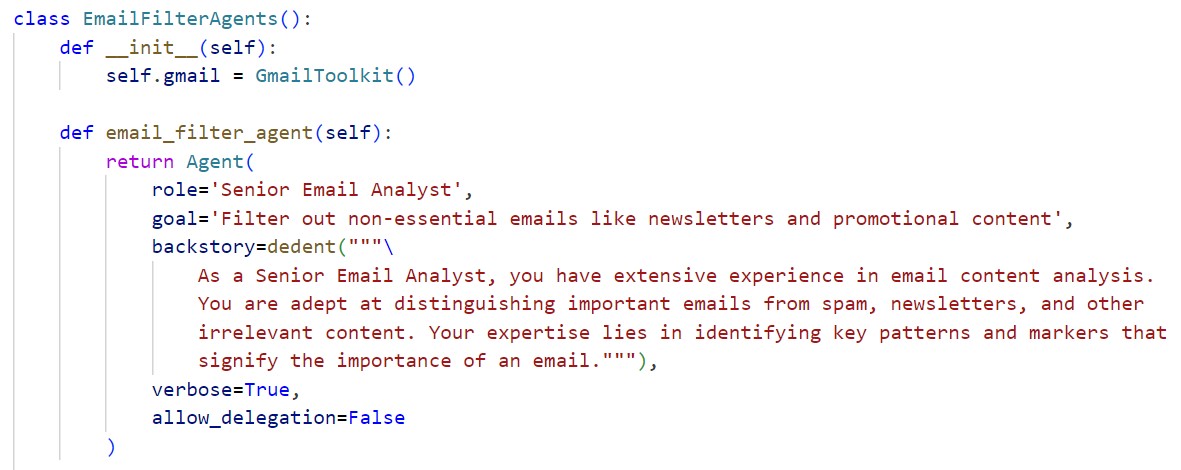}
  \caption{Agent example based on CrewAI.
}
\end{figure}

\begin{figure}[htbp]
  \centering
  \includegraphics[width=\linewidth]{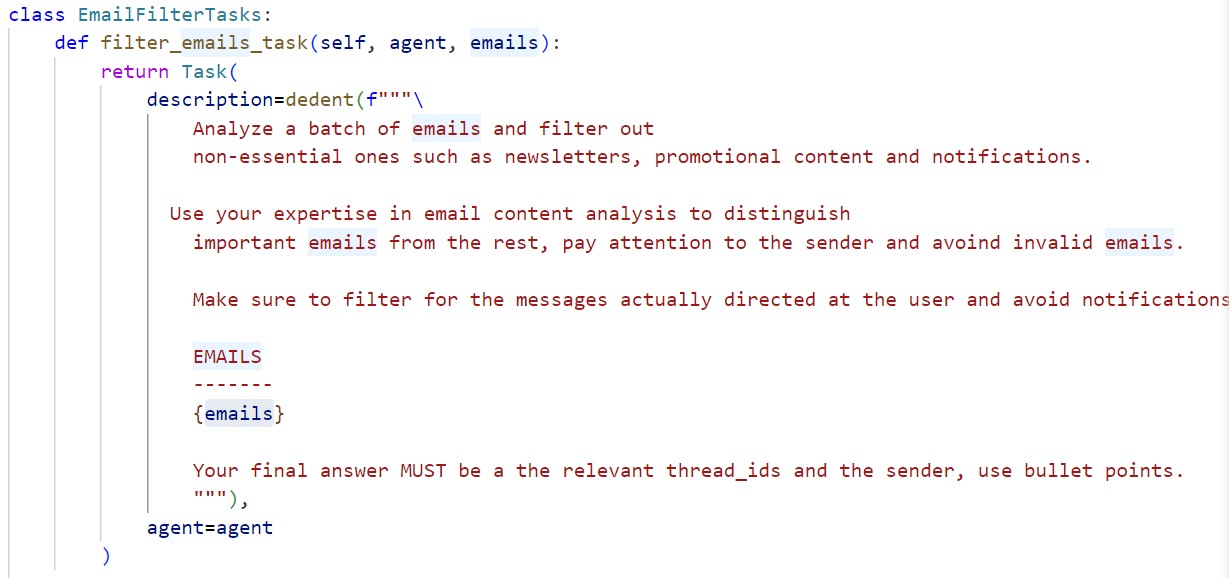}
  \caption{Task example based on CrewAI.
}
\end{figure}

\section{Application Case Practice}

\subsection{Development Environment}
 
In this study, we utilized the following development environment components to apply the LangGraph+CrewAI framework:
\begin{itemize}
\item{Multi-agent: CrewAI, for organizing and coordinating collaborative work among AI agents.}
\item{Graph workflow: LangGraph, used for managing state sharing and process control of graph nodes.}
\item{Workflow tracking: LangSmith, for monitoring and auditing the execution of workflows.}
\item{Embedding technology: ollama embedding model, utilizing embedded vector models to retrieve work order text.}
\item {Vector database: Fasiss is used for storing and retrieving vector data, accelerating data access speed.}
\end{itemize}

\subsection{Application Cases}

As shown in Figure 5, based on the LangGraph+CrewAI framework, we have explored and implemented a multi-agent collaborative application that integrates code generation and code review capabilities. By achieving real-time sharing of status data and feedback mechanisms, the efficiency of code generation has been improved.
\begin{figure}[htbp]
  \centering
  \includegraphics[width=1.0\linewidth]{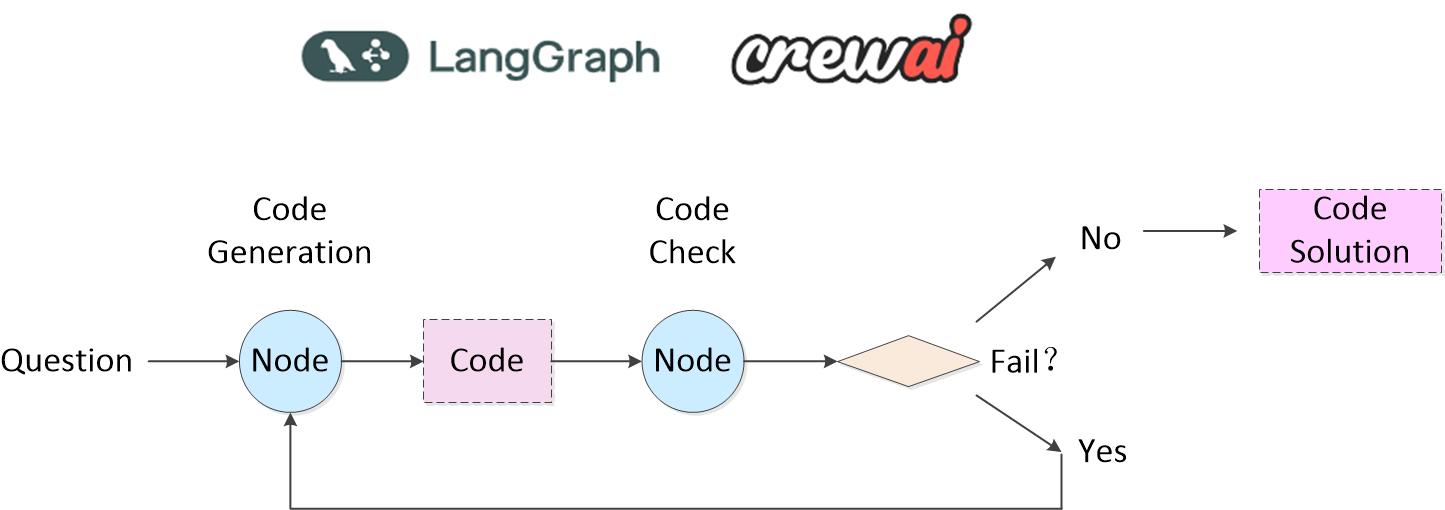}
  \caption{ A code generation case based on LangGraph+CrewAI.}
\end{figure}

As shown in Figure 6, based on the CrewAI+LangGraph framework, we have explored and implemented a case for ticket auditing and forwarding features. By conducting an in-depth analysis of ticket text information, we have built multiple intelligent agents that can more accurately understand the content of the tickets, thereby enhancing the efficiency of ticket processing.
 
\begin{figure}[htbp]
  \centering 
   \includegraphics[width=0.8\linewidth]{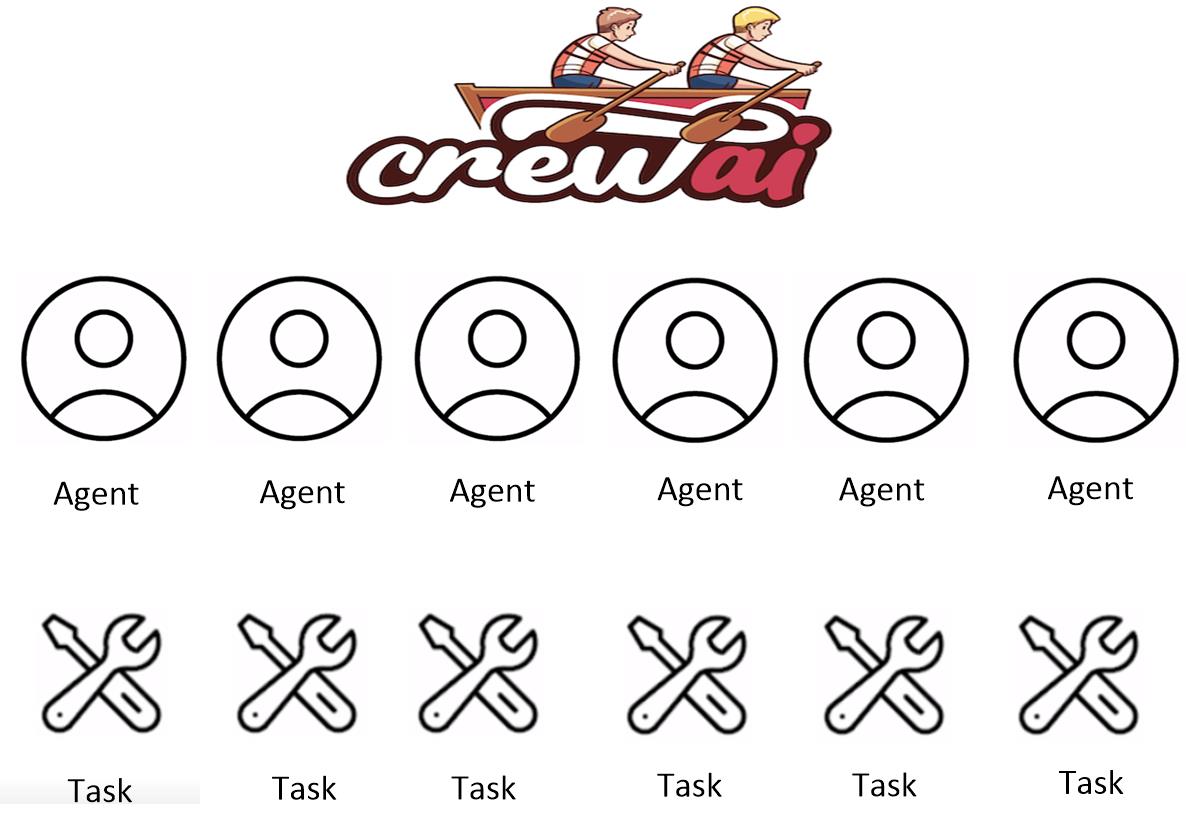}
  \caption{ Case study of automatic processing of work orders based on LangGraph+CrewAI}
\end{figure}

\section{Conclusion}

This article explores the combined application of LangGraph and CrewAI frameworks, demonstrating the powerful capabilities of these two frameworks in building complex multi-agent systems and workflow management. Through case analysis, the LangGraph+CrewAI framework has advantages in task management, inter-agent collaboration, graph workflow implementation, and integration with existing tools. This integrated approach not only improves the efficiency of task execution but also enhances the system's flexibility and scalability through real-time status data sharing and feedback mechanisms. Our research results indicate that the combination of LangGraph and CrewAI provides a powerful toolkit for developing advanced AI applications, especially in scenarios that require handling complex tasks and multi-agent collaboration. By integrating with LangChain, existing independent agents can be easily incorporated into the CrewAI framework, providing developers with a unified platform for building and managing complex AI workflows.

\end{document}